\documentstyle[psfig,floats,epsf,prbbib,aps,pstricks,psfig,pst-plot]{revtex}

\begin{document}

\twocolumn[\hsize\textwidth\columnwidth\hsize\csname
@twocolumnfalse\endcsname

\title{\vspace*{-1cm}\hfill
{\tt Submitted to Phys.~Rev.~Letters}
       \vspace{0.5cm}\\
The hematite ($\alpha$-Fe$_{2}$O$_{3}$) (0001) surface:
evidence for domains of distinct chemistry}

\author{X.-G. Wang, W. Weiss, Sh. K. Shaikhutdinov, M. Ritter, 
M. Petersen, F. Wagner, R. Schl{\"o}gl, and M. Scheffler}
\address{
Fritz-Haber-Institut der Max-Planck-Gesellschaft, Faradayweg 4-6,
D-14195 Berlin-Dahlem, Germany}
\date{\today}
\maketitle
\begin{abstract}
Using spin-density functional theory we investigated various possible
structures of the hematite (0001) surface. Depending on the ambient
oxygen partial pressure, two geometries are found to be particularly
stable under thermal equilibrium: one being terminated by iron and the
other by oxygen. Both exhibit huge surface relaxations 
($-57\%$ for the Fe- and $-79\%$ for the O-termination) with important
consequences for the surface electronic and magnetic properties. With
scanning tunneling microscopy we observe two different surface
terminations coexisting on single crystalline
$\alpha$-Fe$_{2}$O$_{3}$ (0001) films, which were prepared in high
oxygen pressures.
\vspace{0.2cm}

\end{abstract}
\vskip2pc] 
\begin{figure}
\pspicture(5,3.)
\rput[c](3.5,1.0){
\psfig{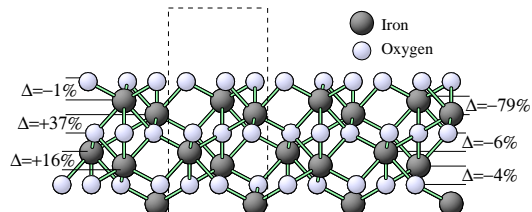}}
\endpspicture
\caption{ Upper half of the 
slab for the unrelaxed O$_{3}$-terminated surface. 
The cross section of the upper half of the
supercell is noted by the dashed rectangle. The $\Delta$ indicate 
the calculated interlayer relaxations in \% relative to an 
unrelaxed surface, i.e., a truncated bulk geometry.
}
\label{fig-slab}
\end{figure}

Although metal-oxide surfaces play a crucial role for several
profitable processes, good quality experimental and theoretical
studies of their atomic structure and electronic properties are
scarce.  For example, $\alpha$-Fe$_{2}$O$_{3}$ appears to be the
active catalytic material for producing styrene, \cite{JWG86} which
was substantiated by recent reactivity studies performed over single
crystalline hematite model catalyst films. \cite{weiss98}  Other
candidate applications are photoelectrodes \cite{LLH92} and non-linear
optics materials \cite{TH95}.  Nevertheless, the surface properties of
$\alpha$-Fe$_{2}$O$_{3}$ are basically unknown, and also for other
metal oxides an understanding is developed only badly. The reason is
the difficult preparation of clean surfaces with defined structures
and stoichiometries, which, as in the case of hematite, can require
high oxygen pressures not suitable in standard ultrahigh vacuum
systems. Furthermore, electron spectroscopy techniques and scanning
tunneling microscopy (STM) are hampered by the insulating nature of
the material. 
We also note that surface-science techniques often do not probe a
thermal equilibrium geometry but a frozen-in metastable state.
Theoretical studies, on the other hand, have to deal
with $3d$ electrons, oxygen with very localized wave functions, a
rather open structure, unusual hybridization of wave functions, huge
atomic relaxations, big super cells, and magnetism. This renders an
{\em ab initio} study of $\alpha$-Fe$_{2}$O$_{3}$ surfaces a most
challenging investigation.  Some theoretical studies of the geometry
of $\alpha$-Fe$_{2}$O$_{3}$ (0001) had been performed using empirical
(classical) potentials \cite{WCM87,EW97}, and Armelao {\em et al.}
\cite{LA95} studied the electronic structure employing a cluster
approach. 
In this paper we report spin-density functional theory
(SDFT) calculations for a slab geometry (see Fig.~\ref{fig-slab}).
We use the generalized gradient approximation (GGA) \cite{JPP92} for the
exchange-correlation functional and the full-potential linearized
augmented plane wave (FP-LAPW) method \cite{PB95,BK96} to solve the Kohn-Sham
equations.
The STM study was performed on a thin hematite film grown epitaxially onto 
a Pt\,(111) substrate.

The identification of thermal equilibrium structures of surfaces is a
prerequisite for an understanding of the endurance, electronic,
magnetic, and chemical properties of the material. For a two-component
material, as Fe$_2$O$_3$, under realistic conditions the surface will
exchange atoms with chemical reservoirs. Therefore we will analyze the
Gibbs free energy with its dependence on 
the chemical potentials of the two components.  For metal oxides, it is
obvious that the O$_2$ partial pressure is the practical handle on
this dependence.  Interestingly, neither experimental nor theoretical
studies have identified the composition and structure of surfaces of
Fe$_2$O$_3$ so far, and it is unknown how the surface structure
depends on the oxygen chemical potential.  In this paper we show that
the surface properties change significantly with O$_2$ pressure and
that under typical oxygen pressure conditions 
two chemically distinct domains are likely to exist.

The $\alpha$-Fe$_{2}$O$_{3}$ crystallizes in the hexagonal corundum structure,
where the primitive bulk unit cell contains six formula units (30 atoms). 
Along the [0001] axis this structure can be viewed as a
stack of Fe and O$_3$ layers:$\cdots$-Fe-Fe-O$_3$-Fe-Fe-O$_3$-$\cdots$.
The (0001) surface is stable and often occurs on naturally grown crystals
\cite{RMC96}. It exhibits an unreconstructed $(1 \times 1)$ surface
with a hexagonal unit cell \cite{Weiss97,TS95}.
In the total-energy calculations we considered the following (likely)
surface terminations with
$(1 \times 1)$ periodicity:
Fe-Fe-O${_3}$-$\cdots$, Fe-O${_3}$-Fe-$\cdots$, O${_3}$-Fe-Fe-$\cdots$, 
O${_2}$-Fe-Fe-$\cdots$,
O${_1}$-Fe-Fe-$\cdots$, where the latter two surfaces correspond to an
oxygen termination with oxygen vacancies. 
For the Fe-O${_3}$-Fe-$\cdots$ terminations we considered
all four possible geometries of the surface Fe, and find that the 
energetically favorable surface is that with the top Fe site occupied 
according to the bulk stacking sequence. Figure~\ref{fig-slab} shows
as an example our slab for the O$_3$-Fe-Fe$-\cdots$ surface study.  We
find that two of the five candidates considered have a particularly
low energy and predict that both should be present under typical
experimental conditions. These are the Fe-O${_3}$-Fe-$\cdots$ and the
O${_3}$-Fe-Fe-$\cdots$ structures. The low energy of the
Fe-O${_3}$-Fe-$\cdots$ surface is consistent with recent findings for
other corundum type materials (Al$_2$O$_3$ \cite{JG92} and Cr$_2$O$_3$
\cite{FR97}), but for Fe$_2$O$_3$ it had not been identified so
far. The other structure which we predict, {\em i.e.}  the termination
by O$_3$, is unexpected in general (also for other corundum type
crystals) because this termination suggests a high surface dipole
moment and thus an electrostatically unfavorable situation. In fact,
our calculations reveal that this argument, which is based on the
understanding of the bulk properties of Fe$_2$O$_3$, is too simple for
describing the surface. Indeed, both low-energy surface structures are
stabilized by a significant surface relaxation which implies that the
surface should be viewed as a ``new material'', not just a truncated
bulk.

For the FP-LAPW calculations we use a kinetic-energy cutoff for the
plane-wave basis of $E_{\rm max}^{\rm wf} = 18$ Ry. This is a rather
high value and makes the calculations very involved.  However, because
of the huge surface relaxations found in the course of this study we
had to use rather small muffin-tin spheres ($R^{\rm MT}_{\rm Fe} =
0.95 $ \AA\,, $R^{\rm MT}_{\rm O} = 0.74 $\AA\,) and therefore a large
value for $E_{\rm max}^{\rm wf}$ was mandatory to ensure good
numerical accuracy. Details of the calculations are presented
elsewhere. 
The bulk as well as the surface calculations were performed with a
hexagonal unit cell and a uniform {\bf k}-point mesh with five points in
the irreducible part of the Brillouin zone.  
All parameters defining the numerical accuracy
of the calculations were carefully tested. 
For the
bulk we obtain the following results (experimental values \cite{LWF80}
in parentheses): $a_{0}$=5.025 (5.035) \AA, $c_{0}$=13.671 (13.747)
\AA, $z$(Fe)=0.357 (0.355) \AA, $x$(O)=0.308 (0.306)~\AA.
Our calculations give an antiferromagnetic ordering with a local
moment of $M = 3.39 \mu_{\rm B}$
(considering the contribution from the muffin tin only). 
The spins are pointing parallel to 
[0001], and in the Fe-Fe double layers 
are aligned parallel. But between neighboring double layers
(separated by an O$_3$ layer) the spins are antiparallel.
Our result is similar to recent bulk calculations using the
augmented spherical wave method \cite{Sandratskii}, but the magnetic moment are smaller 
than the experimental values
$M = 4.6$-$4.9 \mu_{\rm B}$ \cite{JMD71,EK65}. This difference was
blamed on difficulties of the experimental analysis \cite{DDS95,Sandratskii}.
The heat of formation per Fe$_{2}$O$_{3}$ unit is 8.024 
(8.48) \cite{CRC87} eV.
The slabs consist of six O$_3$ layers and ten, twelve, or
fourteen Fe layers, depending on the surface termination to be studied
(see Fig.~\ref{fig-slab} as one example). All atoms of the slab are
allowed to relax and no symmetry restrictions are applied.  The
antiferromagnetic ordering is found to remain up to the surface.

The Gibbs free energy $\Omega$ of the slab at temperature $T$ and
partial pressure $p$ is given by
\begin{equation}
\label{Gibbs}
\Omega=  E^{\rm total} + pV - TS 
- \mu_{{\rm Fe}} N_{{\rm Fe}} - \mu_{{\rm O}} N_{{\rm O}} \quad,
\end{equation}
where $E^{\rm total}$ is the total
energy of the slab,  $\mu_{\rm Fe}$ the chemical potential of iron, 
and $\mu_{\rm O}$ the chemical
potential of oxygen. $N_{\rm Fe}$ and $N_{\rm O}$ are the number of
iron and  oxygen atoms of the supercell. 
For typical pressure and
temperature, the $pV$ and $TS$ terms in Eq.~\ref{Gibbs} can be
neglected.  
The Fe and O chemical potential are not independent,
because they are related to each other by the existence of the
Fe$_2$O$_3$ bulk phase. This gives
\begin{equation}
\label{Gibbs2}
\Omega=  E^{\rm total} - \frac{1}{2}
N_{\rm Fe}\mu_{{\rm Fe}_{2}{\rm O}_{3}({\rm bulk})}+
( \frac{3}{2}N_{\rm Fe} - N_{\rm O}) \mu_{\rm O} \, ,
\end{equation}
where $\mu_{{\rm Fe}_2{\rm O}_3}$(bulk) is the total energy per bulk
Fe$_2$O$_3$ formula unit (with 2 Fe and 3 O atoms).

\begin{figure}

\epsfxsize=7.0cm

\hbox{\hspace{0.2cm}

\epsfbox{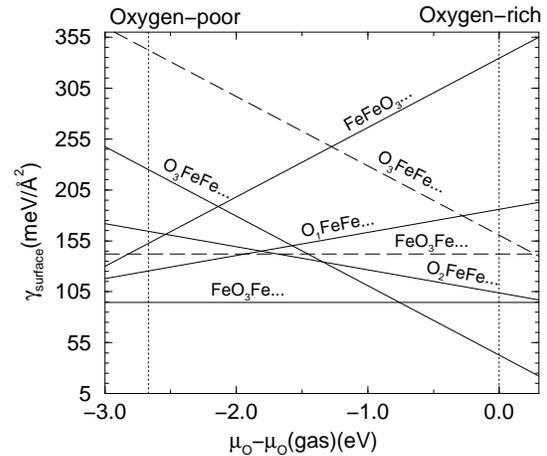}
}
\caption{Surface energies of different Fe$_2$O$_3$(0001) surface
terminations. $\mu_{{\rm O}}(gas)$ is the chemical potential per oxygen atom of 
molecular O$_2$.  The allowed range of
$\mu_{\rm O}$-$\mu_{{\rm O}}(gas)$ is indicated by the vertical dotted lines, where the
left one corresponds to strongly Fe-rich ($i.e.$, oxygen-poor)
conditions, and the right one corresponds to strongly oxygen-rich
conditions ($i.e.$, high oxygen gas pressure). Full lines show results
for relaxed geometries, and dashed lines give for comparison results
for unrelaxed surfaces. }
\label{E-surf}
\end{figure}

In Fig.~\ref{E-surf} we show our results of Eq.~\ref{Gibbs2} for 
various $(1 \times 1)$ geometries, where $\gamma$
is the Gibbs free energy per surface area. The meaningful range of
 $\mu_{\rm O}$-$\mu_{{\rm O}}(gas)$ is limited 
by the conditions that the chemical potential
of Fe has to be smaller than that of an atom of bulk iron, and the
chemical potential of oxygen has to be smaller than that
of an oxygen atom of O$_2$. Otherwise
an iron or oxygen condensate will form at the  surface.
The allowed range of the oxygen chemical potential, 
$\mu_{\rm O}$-$\mu_{{\rm O}}(gas)$, is
marked in Fig.~\ref{E-surf} by the vertical dotted lines.

The results show that the Fe and the O$_3$ terminated surfaces are
particularly stable. The relaxation energy is significant, in particular
for the O$_3$ termination. 
In fact, from these results we predict that 
under typical experimental conditions the surface should consist 
of two domains, with these two structures. 
The O$_3$ termination has not been discussed
before but for the Fe-O$_3$-Fe-$\cdots$ termination semi-empirical
studies had been performed. Our surface energy for the relaxed geometry
is  1.52 $J/m^{2} = 94.58$ meV/\AA \  which is in close agreement with the
semi-empirical studies of of Mackrodt {\em et al.}\cite{WCM87},
who obtained 1.53 $J/m^{2}$, and  Wasserman {\em al.} \cite{EW97},
who obtained  1.65 $J/m^{2}$. Also other corundum-type materials
have a comparable surface energy for the (0001) surface: For 
$\alpha$-Al$_{2}$O$_{3}$ Manassidis {\em et  al.}  \cite{IM93} 
obtained 1.76 $J/m^{2}$, and
Rohr {\em et al.} \cite{FR97} obtained for Cr$_{2}$O$_{3}$ a value of
1.60 $J/m^{2}$.

\begin{table}
\caption{
Interlayer relaxations at the Fe-O$_3$-$\cdots$ and O$_3$-Fe-$\cdots$ surface
relative to the corresponding bulk spacings.}
\label{table-Fe}
\begin{tabular}{c|ccc|c|c}
           & \multicolumn{3}{c|}{Fe-O$_3$-\dots}  &  & O$_3$-Fe-\dots \\
interlayer & present &   Ref.~\onlinecite{WCM87} & Ref.\onlinecite{EW97} & interlayer & present \\
\tableline
Fe-O$_3$ \quad 1-2     &$-$57$\%$ &  +1$\%$   & $-$49$\%$ & & \\
O$_3$-Fe \quad 2-3     &   +7$\%$ &  +5$\%$   & --3$\%$  & 1-2 & $-$1 \% \\
Fe-Fe \quad 3-4        &$-$33$\%$ &$-$47$\%$  & --41$\%$ & 2-3 & $-$79 \% \\
Fe-O$_3$ \quad 4-5     &  +15$\%$ &  +20$\%$  &  +21$\%$ & 3-4 & $+$37 \% \\
O$_3$-Fe  \quad 5-6    &   +5$\%$ &   +3$\%$  &   --     & 4-5 & $-$6 \% \\
Fe-Fe  \quad 6-7       & $-$3$\%$ &   +2$\%$  &   --     & 5-6 & $+$16 \% \\
Fe-O$_3$  \quad 7-8    &   +1$\%$ &   --      &   --     & 6-7 & $-$4 \% \\
O$_3$-Fe  \quad 8-9    &   +4$\%$ &   --      &   --     & 7-8 & $+$4 \%  \\
\end{tabular}
\end{table}

Table~\ref{table-Fe} collects the calculated surface relaxations for the
Fe-terminated surface. We obtain a relaxation pattern which alternates
when going away form the surface.  For the Fe terminated surface, the
comparison with previous (semiempirical) studies reveals that a model
using static ions \cite{WCM87} is not appropriate, but when some
polarization is allowed \cite{EW97}, the system is apparently well
described. In fact, our calculation shows that at the surface the
covalent contribution to bonding is enhanced, which is mainly due to a
hybridization of O $2p$ and Fe $3d$ orbitals.  
The workfunction of the Fe terminated surface is calculated as 4.3~eV 
(before relaxation it is only 3.1~eV).
Recently, in a LEED
analysis for Cr$_{2}$O$_{3}$ Rohr {\em et al.} \cite{FR97} determined
a 60~\% reduction for the first Cr-O$_3$ interlayer distance, which is
in surprisingly close agreement with our result for Fe$_{2}$O$_{3}$
(we obtain 57~\%).  A detailed list with the geometry data 
which we obtained
will be given elsewhere.
Here we only note that the O$_3$ triangles of the first oxygen layer,
$i.e.$ the second layer from the surface,
exhibit a clock-wise rotation by 2$^\circ$ 
without breaking the
C$_3$ surface symmetry.
No LEED analysis for the Fe-terminated (0001) surface of 
Fe$_{2}$O$_{3}$ exists so far.

The O$_3$ terminated surface (see Figs.~\ref{fig-slab} and \ref{E-surf})
is indeed a very unexpected system and only stabilized by
huge and unusual  relaxations which are collected in Table~\ref{table-Fe}.
The top layer O$_3$ triangles undergo a significant rotation (by
10$^\circ)$ without breaking the C$_3$ surface symmetry.
The oxygen atoms remain nearly planar. Similarly to the Fe-terminated
surface, a very notable feature is the contraction of the first
subsurface Fe-Fe interlayer spacing, followed by an expansion of the
next Fe-O$_3$ spacing. The huge
interlayer relaxations and the O$_3$ rotational reconstruction provide
the important contribution to the decrease of the surface energy and
stabilize the O$_{3}$-terminated surface.  As mentioned above, the
stability of this surface cannot be understood in terms of a simple
ionic model, in which the surface oxygen atoms where negatively
charged.  Indeed, according to our calculations the ionic character of
the O atoms is decreased at the O$_{3}$-terminated surface, and the
covalent interaction between O $2p$ and Fe $3d$ states is enhanced
significantly. This is also reflected in a noticeable {\em decrease} of the 
work function upon relaxation (from 8.3~eV to 7.6~eV).

The surface relaxations and change in the nature of bonding
at the surface goes hand in hand with 
a dramatic change in the magnetic properties. 
Here, we discuss the integral over the magnetization density
considering the contribution from inside the muffin-tin spheres.
Therefore the direct values should not be taken too literately. 
But differences in moments between different
layers and between the relaxed and unrelaxed situation are indeed meaningful.  
For the Fe-terminated surface we find that the magnetic moment of the
topmost Fe layer is reduced against the bulk value by about
 7~\% (namely $\Delta M = -0.24 \mu_{\rm B}$). 

The electronic and magnetic properties of the O$_{3}$-terminated
surface are unprecedented. We find that the subsurface iron layers
change their character significantly. The local magnetic moments of
the first and second Fe layer are reduced to less than 50 \% of the
bulk values ($\Delta M$ = -1.79 and -1.75 $\mu_{\rm B}$ respectively), 
and again, a significant fraction of this reduction only
occurs upon surface relaxation and reconstruction.  Interestingly,
also the top O$_3$ layer now attains a noticeable
magnetic moment (M=0.20 $\mu_{\rm B}$ per atom). In the bulk, 
the magnetic moment of oxygen atoms is zero. 
Our calculations show that the surface states for the
O$_{3}$-terminated surface are of Fe 3$d$ character.
They are partially occupied by 
the Fe 3$d$ electrons formerly with different spin,
which results in the decrease of the local magnetic moments of the iron
atoms. 
Our calculations also show that the surface state electrons of the
iron atoms reaching through the topmost O layer 
may noticeably affect the surface reactivity.

The STM experiments were performed in an ultrahigh vacuum system
described in detail elsewhere~\cite{weiss98b}.
It is equipped with a
sample transfer mechanism and a separate
preparation chamber for performing high pressure oxidation
treatments. Single crystalline $\alpha$-Fe$_2$O$_3$(0001) films were
grown onto a clean Pt(111) substrate surface. In a first step about 10
nm thick Fe$_3$O$_4$(111) magnetite films were prepared by repeated
deposition of iron and subsequent oxidation for 2 min at temperatures
around 950 K in 10$^{-6}$ mbar oxygen partial pressure\cite{AB94}.
Then a final
oxidation at T=1100 K in 10$^{-3}$ mbar oxygen partial pressure was
performed for 10 min.  The structure formed under these conditions was
quenched by cooling down the sample to room temperature in the oxygen
atmosphere. Then the oxygen was pumped off and the sample was
transfered back into the analysis chamber and immediately studied by
STM. These films exhibited no Auger
contamination signals and sharp
(1$\times$1) LEED patterns~\cite{Weiss97}.
X-ray photoemission spectroscopy revealed only Fe$^{3+}$ and no Fe$^{2+}$
species, indicating the
formation of a single phase hematite film \cite{TS95}. 

\begin{figure}[t]
\begin{center}
\pspicture(5,5.5)
\rput[c](1.0,2.8){
\psfig{figure=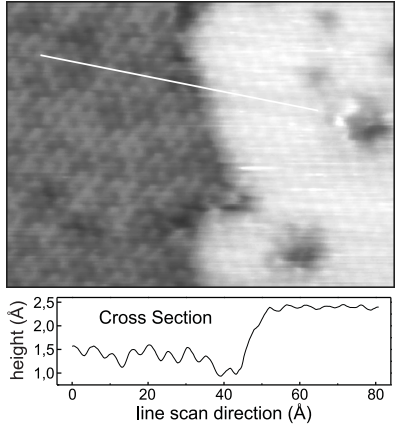,height=6cm}}
\endpspicture
\end{center}
\caption[]{Constant current image of an $\alpha$-Fe$_2$O$_3$~(0001)
surface prepared in 10$^{-3}$ mbar oxygen partial pressure over an
area of 100$\times$80~\AA$^{2}${} ($U$=+1.3~V, $I$~=~1.3~nA). 
Below a cross section along the
white line in the image is shown. Two different surface terminations
can be seen.}
\label{STM}
\end{figure}
Figure~\ref{STM} shows an atomic resolution STM image of the
$\alpha$-Fe$_2$O$_3$(0001) surface. A dark region and a bright region
can be seen.  They both exhibit hexagonal lattices formed by atomic
protrusions with a periodicity of 5~\AA, which corresponds to the
interatomic distance within the (0001) Fe-planes of the
$\alpha$-Fe$_2$O$_3$ structure. Below a cross section along the white
line in the STM image is shown, which displays the 5~\AA{} periodicity
and the two regions separated by a step about 1~\AA{} high. This
step height considerably deviates from the distance between two
equivalent (0001) surface terminations of $\alpha$-Fe$_2$O$_3$, which
are separated by a monoatomic step with a height of 2.28~\AA. We
always measure monoatomic steps or multiples of them between
equivalent regions on the hematite surface.  Therefore the two regions
in Fig.~\ref{STM} correspond to two different surface terminations
that coexist on the $\alpha$-Fe$_2$O$_3$(0001) surface. This is
further substantiated by their different corrugation amplitudes
depicted in the cross section plot in Fig.~\ref{STM}, which are 0.1 and
0.2-0.3~\AA. Based on the apparent topographic height difference
measured with the STM, we interpret the bright region as the
Fe-terminated and the dark region as the O-terminated surface.
About 20~\% of the entire sample surface are covered by the bright
termination as determined by numerous large-area scans.  They form
islands preferentially located near step edges on top of the dark
termination, with sizes ranging between 20 and 200~\AA. We also
confirmed that their formation depends on the oxygen partial pressure
during the preparation (compose the discussion of Fig.~\ref{E-surf}
above). Measurements of STM spectroscopy, work function etc., and a
detailed analysis of the dependence on oxygen pressure are in
progress.

In summary, we have presented a detailed {\em ab initio} FP-LAPW study
of the stoichiometry and structural relaxations of
the $\alpha$-Fe$_{2}$O$_{3}$ (hematite) (0001) surface. The calculations
predict two kinds of surfaces (thermal equilibrium with (1$\times$1)
symmetry), which depend on the growth conditions.
We find a large inwards relaxation of the first layer for the
Fe-terminated surface and a huge contraction of the interlayer spacing
between the Fe subsurface layers for the O$_{3}$-terminated surface.
In addition to the relaxations along the [0001] direction, the O
layers of both surfaces have a plane rotational reconstruction. To our
knowledge, this is the first finding for such kind of special
relaxations and plane reconstruction of oxygen layers at a metal-oxide 
surface. We also predict an unusual electronic structure of the
O$_3$-terminated surface with noticeable presence of states from the
subsurface Fe layer. This also results in a magnetic polarization of
the oxygen. With STM we confirm the existence of two different
surface terminations coexisting on single crystalline hematite (0001)
films, which have been prepared under a high oxygen pressure of
10$^{-3}$~mbar.


\begin{references}
\bibitem{JWG86} T. Hirano, Appl. Catalysis {\bf 26}, 65 (1986).

\bibitem{weiss98} W. Weiss {\em et al.},
Catal. Lett. 52, 215 (1998).

\bibitem{LLH92} L. L. Hu, T. Yoko, {\em et al.},
 Thin Solid Films {\bf 219}, 18 (1992).

\bibitem{TH95} T. Hashimoto, {\em et al.},
J. Ceram. Soc. Japan {\bf 101}, 64 (1993).

\bibitem{WCM87} W. C. Mackrodt. {\em et al.},
J. Cryst. Growth, {\bf 80}, 441 (1987). 

\bibitem{EW97} E. Wasserman, {\em et al.},
Surf. Sci. {\bf 385}, 217 (1997).

\bibitem{LA95} L. Armelao, {\em et al.},  
J. Phys.: Cond. Matt. {\bf 7}, L299 (1995).

\bibitem{JPP92} J. P. Perdew, {\em et al.},
Phys. Rev. B {\bf 46},  6671 (1992). 

\bibitem{PB95} P. Blaha, {\em et al.},
Comput. Phys. Commun. {\bf59}, 399 (1990).

\bibitem{BK96} B. Kohler, {\em et al.},
 Comput. Phys. Commun. {\bf 94}, 31 (1996).

\bibitem{RMC96} R. M. Cornell, {\em et al.}, (ed.), $The$ $Iron$ $Oxides$,
VCH, Verlagsgesellschaft, (1996).
  
\bibitem{Weiss97} W. Weiss, Surf. Sci. {\bf 377-379}, 943 (1997).

\bibitem{TS95} Th. Schedel-Niedrig, {\em et al.},
 Phys. Rev. B {\bf 52}, 17449 (1995).

\bibitem{JG92} J. Guo, {\em et al.},
Phys. Rev. B {\bf 45},  13647 (1992).

\bibitem{FR97} F. Rohr, {\em et al.},
Surface Science {\bf 372}, L291 (1997).

\bibitem{LWF80} L. W. Finger,
{\em et al.}, J. Appl. Phys. {\bf 51}, 5362 (1980).

\bibitem{Sandratskii} L. M. Sandratskii, {\em et al.},
J. Phys.: Condens. Matter {\bf 8}, 983 (1996).

\bibitem{JMD71} J.M.D. Coey {\em et al.},
J. Phys. C {\bf 4}, 2386 (1971).

\bibitem{EK65} E. Kren,{\em et al.},
Phys. Lett. {\bf 19}, 103 (1965).

\bibitem{DDS95} D. D. Sarma  {\em et al.},
Phys. Rev. Lett. {\bf 75}, 1126 (1995).

\bibitem{CRC87} CRC Handbook of Chemistry and Physics, edited by 
R. C. Weast, (CRC Press, Boca Raton, Florida, 1987), 67th ed. 

\bibitem{IM93} I. Manassidis, {\em et al.},
 Surf. Sci. Lett. {\bf 285}, L517 (1993).

\bibitem{weiss98b} W. Weiss, {\em et al.},
J. Vac. Sci. \& Technol. A16(1), 21 (1998).

\bibitem{AB94} W. Weiss, {\em et al.}, 
Phys. Rev. Lett. {\bf 71}, 1848 (1993).

\end{references}
\end{document}